\documentclass[prl,amsmath,amssymb ,twocolumn]{revtex4}

\usepackage[dvips]{graphicx}
\usepackage{times,color}

\usepackage{amssymb,amsfonts,amsmath,multirow}

\newcommand{\beq}{\begin{equation}}
\newcommand{\eeq}{\end{equation}}
\newcommand{\beqa}{\begin{eqnarray}}
\newcommand{\eeqa}{\end{eqnarray}}

\begin{document}

\title{Small regulatory RNAs may sharpen spatial expression patterns}

\author{Erel Levine\footnote{These authors contributed equally to this work.}, Peter McHale\footnotemark[\value{footnote}], and Herbert Levine}
\affiliation{Center for Theoretical Biological Physics,
University of California at San Diego, La Jolla, CA 92093}

\begin{abstract}
The precise layout of gene expression patterns is a crucial step in development. 
Formation of a sharp boundary between high and low expression domains requires a genetic 
mechanism which is both sensitive and robust to fluctuations, a demand that may not be easily 
achieved by morphogens alone. 
Recently it has been demonstrated that small RNAs (and, in particular, microRNAs) play many roles in embryonic development. 
While some RNAs are essential for embryogenesis, others are limited to fine-tuning 
a predetermined gene expression pattern. 
Here we explore the possibility that small RNAs participate in sharpening a gene expression profile that was crudely established by a morphogen. To this end we study a model where small RNAs
interact with a target gene and diffusively move from cell to cell. 
Though diffusion generally smears spatial expression patterns,
we find that intercellular mobility of small RNAs  is actually critical in sharpening the interface between target expression domains in a robust manner.
We discuss the applicability of our results, as examples, to the case of leaf polarity establishment in maize and Hox patterning in the early {\it Drosophila} embryo. 
Our findings point out the functional significance of some mechanistic properties, such as mobility of small RNAs and the irreversibility of their interactions. These properties are yet to be established directly for most classes of small RNAs. An indirect yet simple experimental test of the proposed mechanism is suggested in some detail.
\end{abstract}

\maketitle





\section{Introduction}

Morphogenesis proceeds by sequential divisions of
a developing embryo into
{domains}, each expressing
a distinct set of genes. Each combination of genes
is associated with a particular cell identity.
At advanced stages of development,
{most genes that define cell identity are
either highly expressed (`on') or strongly inhibited (`off') in a given cell.}
{For example, two adjacent domains may be differentiated by a high expression of some
genes in one, and low expression in the other. }
In such cases {it is important} that cells of the two populations do
not intermix.  Moreover, the number of cells that
 show intermediate levels of expression, typically
found at the interface between the two sets, should be kept to a minimum.
These demands are
necessary in order to unambiguously define the identity of each cell.
A {spatial gene expression pattern} that obeys these demands is said to exhibit a {\em sharp interface}.

A crucial step in
setting the interfaces of gene expression patterns
is often the establishment of a
concentration gradient of molecules called morphogens.
Some morphogens are transcription factors which regulate gene expression
directly \cite{NussleinVolhard1988a,NussleinVolhard1988b}.
Others are ligands that bind cell-surface receptors
signaling the activation of target expression \cite{OConnor2006}.
Since morphogens
act in a concentration-dependent manner,
a morphogen gradient may be transformed into a
gradient of its target messenger RNA (mRNA).

In principle, a single morphogen interacting cooperatively with its target enhancer can
generate a sharp interface in the target transcription profile, by modulating the rate of its mRNA transcription as a function of the nuclear spatial coordinate \cite{Janssens2006}.
This may be done e.g. by cooperative binding to a receptor or to a promoter
\cite{Lebrecht2005} or by zero-order ultrasensitivity \cite{Melen2005}.
As an example, in fruit-fly early embryonic development
Hunchback transcription depends on the cooperative binding of
about $5$ Bicoid molecules \cite{Gregor2007b}.
{An obvious limitation in this mechanism is the need for
large cooperativity factors or cascades of reactions,
which make it prone to fluctuations and
slow to adapt \cite{Berg2000,Thattai2002,Gregor2007b,Howard2007}.}
Recently, the role of small regulatory RNAs in establishing developmental patterning has been observed in plants \cite{Juarez04,Kidner04, Nogueira07} and animals \cite{Aboobaker2005}. In particular, it has been suggested that microRNAs (miRNAs) confer accuracy to developmental gene expression programs \cite{Stark2005}. This raises the possibility that small RNAs aid  morphogen gradients in establishing sharp interfaces between `on' and `off' target-gene expression.

In this study we formulate a mathematical model in which small regulatory RNA 
help morphogens to determine cell identity by
sharpening morphogen-induced expression patterns.
For specificity we assume here that the small RNA belongs to the miRNA family, and consider other classes of small RNA in the Discussion. 
miRNAs constitute a major class of gene
regulators that silence their targets by
binding to target mRNAs.
In metazoans primary miRNA transcripts are transcribed and
then processed
both inside and outside of the nucleus to form
mature transcripts approximately 21 nt in length that
are then loaded into the RNA-induced silencing complex (RISC)
\cite{Kloosterman2006}.
They are found in plants \cite{Baulcombe2004}
and animals \cite{Bartel2004}, including human
\cite{Lewis2005}, and are
predicted to target a large fraction of all
animal protein-coding genes \cite{Brennecke2005,Grun2005,Lewis2005}.
%
In plants miRNAs are known to affect morphology
\cite{Chen2004,Juarez04,Kidner04} implying that
they play an important role in determining cell identity.
This is underscored by the fact that the spatiotemporal
accumulation of miRNAs is under tight control in plants \cite{Valoczi2006},
fly \cite{Aboobaker2005,Biemar2005} and zebrafish \cite{Giraldez04072006}.

{Our model is constructed in one spatial dimension, namely along one spatial axis.
Domains of gene expression are laid out along this axis, and we assume no significant variance along other, perpendicular axes.  }
Two key ingredients of the model are a {strong} interaction
between miRNA and mRNA
and intercellular mobility of the miRNA.
Within this framework miRNAs generate a sharp interface between those cells
expressing high levels of the target mRNA and those expressing
negligible levels of mRNA.
We use physical arguments
to understand the range of parameters where this sharpening occurs.
Our model predicts that the spatial position of the interface is precisely determined: 
mobile miRNAs spatially average individual cellular fluctuations without 
compromising the interface sharpness. We use computer simulations to show that 
this is also true even with low copy numbers. 
A consequence of our model is that a local {change} to
the transcription profiles can induce a nonlocal effect on the mRNA
concentration profile; we outline an experiment to detect this
nonlocal property.
We also consider the special case where
the miRNA and target transcription profiles coincide \cite{Mansfield2004}.
We show that miRNAs preserve their sharpening function provided
mobile miRNAs change irreversibly to an immobile
state before interacting with their targets.
By separating mobility and interaction in this way,
the target is rescued from miRNA-mediated silencing in selected
cells thereby generating a sharp expression pattern. Finally we consider possible applications of these ideas in plants and fruit fly.


\section{Model and results}

\subsection{Formulation of model from available biological evidence}

Our theory comprises three central elements.
First, miRNA and its target are taken to be transcribed
in a space-dependent manner. Second, {we assume that the interaction between
miRNA and target irreversibly disable the target mRNA from being translated into proteins; this for example may be done by promoting the degradation of the target. Furthermore, the miRNA molecule itself may be consumed during this interaction.   }
Last, we allow for the possibility that miRNAs move between
cells.
Before defining the model, let us review the available data regarding each
of these processes.

microRNAs and their targets are often expressed in a coordinated manner \cite{Hornstein2006}.
{We consider two classes.}
In the first, the regulatory network is designed to express the miRNA and its
targets in a mutually exclusive fashion. For example,
the expression patterns  of
the miRNA miR-196 and its target Hoxb8
are largely non-overlapping in mouse \cite{Mansfield2004}
and chick \cite{Hornstein2005}.
Similarly, the nascent transcripts of Ubx,
still attached to the DNA, are expressed in a stripe near
the center of the early embryo, while nascent {transcripts of its
regulator,} iab-4{, are} simultaneously
observed in nuclei posterior to this domain
\cite{Ronshaugen2005}.
A recent large-scale study in fly showed that miRNAs and their
target genes are preferentially expressed in neighboring tissues
\cite{Stark2005}. Likewise, in mouse \cite{Farh12162005}
and in human \cite{Sood2006}
predicted miRNA targets were found at lower levels
in tissue expressing the cognate miRNA than in other tissues.

{Cases in the second class }
are characterized by co-transcription of the miRNA and its targets \cite{Hornstein2006}, as occurs for example in mouse cardiogenesis \cite{Zhao2005} and human cell proliferation \cite{ODonnell2005}.
Likewise, microRNAs encoded within the Hox gene cluster may target co-expressed Hox genes \cite{Mansfield2004}.

Our model assumes that the synthesis rate of the miRNA and its target are {\em smoothly} varying along a spatial axis, $x$. 
This, for example, may be the result of  a common morphogen regulating (either directly or indirectly) the two species.
The transcription profiles $\alpha_\mu(x)$ and $\alpha_m(x)$ of the miRNA and its target, resp., may be anti-correlated (the first class above)
or highly correlated (the second class).

The detailed interaction between miRNAs and their targets
is  currently a topic of intense investigation \cite{ValenciaSanchez2006,Pillai2007}.
miRNAs induce the formation of a ribonucleoprotein complex (RISC).
Targeting of a specific mRNA by a RISC is done via (often imperfect) base-pair complementarity to the miRNA \cite{Bartel2004}.
Upon binding, protein synthesis is suppressed by either translation inhibition or mRNA destabilization \cite{ValenciaSanchez2006,Pillai2007}. While it is likely that miRNA can go through a few cycles of mRNA binding \cite{Zamore2004}, the increased endonucleolytic activity conferred by the miRNA  makes it plausible that the miRNA is sometimes degraded in the process. In addition, evidence suggests that mRNAs which are translationally repressed by miRNA may be co-localized to cytoplasmic foci {such as processing-bodies \cite{Liu2005, ValenciaSanchez2006,Pillai2007}, {which are enriched with endonucleases,} or stress granules \cite{Leung2006}.} Taken together, these facts make it improbable that miRNAs act in a fully catalytic manner.

A pair of mRNA-miRNA reactions that describe a spectrum of plausible
scenarios is
\beq
m+\mu \stackrel{\theta k}{\longrightarrow} \mu,
\,\,\,  m+ \mu \stackrel{k}{\longrightarrow} \emptyset, 
\label{basicModel}
\eeq
where $m$ represents the mRNA concentration and $\mu$ represents that
of the miRNA. Here, $\theta$ is the average number of targets degraded
by a given miRNA before it is itself lost in the process.
These reactions may be realized in different ways.
For example, the two species may reversibly form a
complex that is then subject to degradation.
Another possibility is that  the two species irreversibly associate to form an inert complex.
Furthermore, the reaction between the species may be reversible, as long as
the typical dissociation time is much longer than the
relevant biological time scale.
One way in which the cell may control {the dissociation time} is
by regulating exit of the RNA pairs from P-bodies \cite{Levine2007_miRNA}.

Can miRNAs move from cell to cell?
siRNAs, another important class of small
RNAs, are known to illicit non cell autonomous RNA silencing
in plants, worms, fly and possibly mouse (reviewed in
\cite{Voinnet2005}). This may also be the case for trans-acting siRNA \cite{Nogueira07}.
Evidence in favor of intercellular mobility of miRNA is found in
pumpkin \cite{Yoo2004}. There, miRNAs have been found in the phloem sap which is transported
throughout the plant by phloem tissue.
In animals, many small RNAs, including many miRNAs, were found in exosomes from mouse and human mast cell lines, which can be delivered between cells \cite{Valadi07}. 


In our model we {consider}
the possibility that miRNA migrate from cell to cell.
Mobility of the miRNA species is likely to rely on active
export from the cell followed by import to neighboring cells,
or perhaps on direct transport between cells.
On the tissue scale, these transport processes are
expected to result in effective diffusion. We therefore ignore the
small-scale transport processes and model miRNA mobility as pure diffusion.

%

Finally, we combine these processes into a steady-state mean-field model given by
\begin{subequations}
\label{annModelEqs}
\beq
0  =  \alpha_m - \beta_m m - k m \mu \label{annModelEq1}
\eeq
\beq
0  =  \alpha_\mu - \beta_\mu \mu - k m \mu + D \mu''.\label{annModelEq2}
\eeq
\end{subequations}
The $\beta$ terms describe {autonomous degradation
(i.e., by processes independent of the other RNA species)}
and the $k$ term
{describes coupled degradation of both RNA species.}
Note that the case $\theta\neq0$ ({where miRNA may go through multiple rounds of interactions with target mRNAs}) can also
be brought into this form by {rescaling} \eqref{annModelEq1}.
Mobility of the miRNA is described by an effective diffusion constant $D$. 
The spatial coordinate $x$ measures distance along
one dimension of a tissue.
All our numerical results shall be presented in units of the
{tissue size}, i.e. $0\leq x \leq 1$.

\subsection{microRNAs may sharpen domain boundaries.}
\label{mobile_miRNA_section}

As described above, a desired {target protein profile  comprises a
domain of cells which express this protein abundantly, adjacent} to
a domain of cells where this protein does not accumulate.
Furthermore, one requires that the number of cells with intermediate
expression levels lying in between the two domains be
minimized---this is our definition of a sharp interface. In this
section we discuss one scenario where the mutual consumption of a
diffusive miRNA and its target {leads} to such a sharp interface in
the mRNA profile. 

{We assume that some morphogen controls the transcription rate of
the target. The transcription profile---{namely,} the transcription
rate as a function of the spatial coordinate of the nucleus---is
laid down as a smooth gradient, falling from one end 
of the developing tissue (which, for convenience, will be called `left') towards the other end (`right'). Motivated by a
recent study that showed that miRNA and their targets are
preferentially expressed in neighboring tissues \cite{Stark2005},}
we  focus on the scenario where the miRNA transcription is
controlled in a fashion opposite to that of the mRNA: {miRNA
transcription is peaked at one end of the tissue (where the mRNA
transcription rate is minimal), and decreases towards the other end
(where the mRNA transcription rate is maximal)}.  The kind of {\em
`mutually exclusive'} transcription profiles we have in mind is
depicted in {Fig.~\ref{thresholdFigure}A}. In this figure, and
hereafter, we denote the mRNA transcription profile by $\alpha_m(x)$
and the miRNA transcription profile by $\alpha_\mu(x)$, explicitly
noting their dependence on the spatial coordinate $x$. {We note
that, in the absence of miRNA-target interaction and of miRNA
diffusion, the concentration profiles of mRNA and miRNA simply
follow their transcription profiles, Fig.~\ref{thresholdFigure}A and~D.
Each is rather smeared and overlaps the other near the centre of the
tissue.}

{Before studying the full model,} it is instructive to consider first the {interacting} system in the absence of
miRNA diffusion. {In the context of mutually exclusive transcription profiles, we expect that {each cell} would be dominated by one RNA species (either the miRNA or the target mRNA), which we will call
the {\em majority species}, and be depleted of the other, {\em minority species}. {In other words, we are making} the critical assumption that the decay of the minority species {\em in each cell} is governed by the interaction with the other species (rather than by its {autonomous} degradation). This assumption can be made quantitative in terms of the model parameters; see Eq.~\ref{criticalAssumption} of the Supporting Information. Under this assumption, which {we refer to as} the {\em strong interaction limit}, it is straightforward to show (Eq.~\ref{thresholdResponse} of the Supporting Information) that the density of the majority species in each cell is proportional to the {\em difference} between the two transcription rates in that cell, whereas the minority species is essentially absent. Consequently, in the context of mutually exclusive transcription profiles, the mRNA level becomes vanishingly small in any cell for which $\alpha_m(x)<\alpha_\mu(x)$, namely every cell to the {right} of the point where the two {transcription} profiles are equal. {The} concentration profiles of the two RNA species are shown in Figs.~\ref{thresholdFigure}B and~E,
{where one can see that the mRNA and miRNA spatial expression domains are now complementary and more sharply defined.}

{The {\em threshold response} {that arises} when both RNA species
do not diffuse from cell to cell} provides insurance against
the possibility that the
mRNA transcription profile is not as step-like as is required
for {unambiguous} cell differentiation. In other words, miRNA
regulation acts as a failsafe mechanism whereby incorrectly
transcribed low-abundance transcripts
in the region $\alpha_m(x)<\alpha_\mu(x)$ are silenced,
while correctly transcribed high-abundance transcripts
in the region $\alpha_m(x)>\alpha_\mu(x)$
are only mildly affected \cite{Stark2005,Hornstein2006}.
This threshold response in the
target profile has been observed in the context of small
RNAs---another class of post-transcriptional regulators---in
bacteria \cite{Levine2007_sRNA}.

\begin{figure*}
\begin{center}
\includegraphics{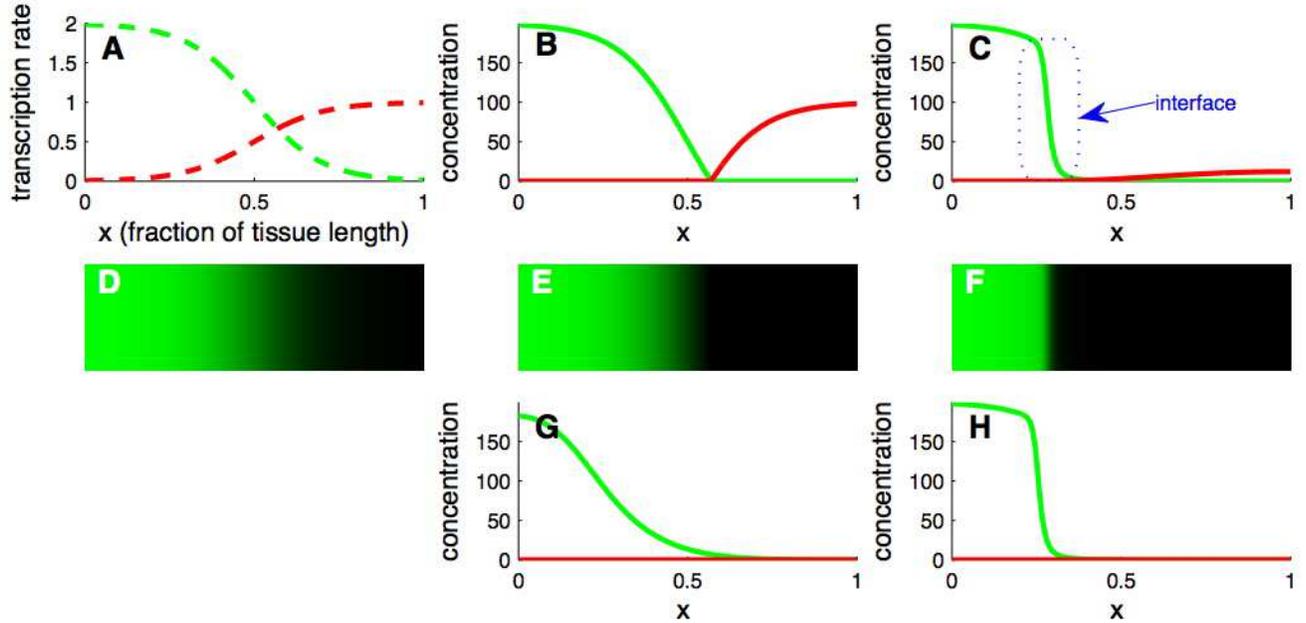}
\end{center}
\caption{{Sharpening the target expression pattern.
{\bf A.} Transcription profiles of a miRNA (red) and its target (green). Functional expressions and parameter values are given in the Supporting Information, unless otherwise noted. 
{\bf B.} Steady state concentration of target-mRNA (green) and cognate miRNA (red). Regulation by immobile miRNAs (where the diffusion constant is $D=0$) removes residual mRNAs from the right domain, creating a sharp boundary near the center of the tissue.
{\bf C.} Mobility of miRNA ($D=0.01$) further sharpens this boundary by inducing an interface between domains of gene expression.
{\bf D-F.} Steady-state concentration of mRNA (green level) for a two-dimensional generalization of the model, where diffusion occurs equally in both directions and transcription rates do not vary along the vertical axis. Parameter values correspond, respectively, to panels A-C. 
{\bf G.} Strong miRNA diffusion ($D$ increased 100-fold) smears the interface between the domains, but does not affect the interface position.
{\bf H.} The interface structure is unaffected by a increasing both the miRNA-mRNA interaction $k$ and the diffusion constant $D$ 100-fold.}
\label{thresholdFigure}}
\end{figure*}


{We now return to our full model, which allows for diffusion of the miRNA.
To simplify the analysis let us keep the strong-interaction limit, described above (and in Eq.~\ref{criticalAssumption} of the
Supporting Information). In general,} one expects that diffusion
makes the miRNA profile more homogeneous,
and this is confirmed by exact {numerical} solution of the model,
as shown in Fig.~\ref{thresholdFigure}C.
{Surprisingly however,} the mRNA profile does not become smoother. {In fact, Figs.~\ref{thresholdFigure}C and F show
that this profile {actually develops}}
a sharper drop from high to low mRNA levels than there was in the
absence of diffusion.
More specifically,
miRNA diffusion creates an interface between
high and negligible target expression.
Increasing diffusion moves the interface deeper into
the mRNA-rich region and thereby accentuates the drop in
mRNA level across the interface.
Although some miRNA diffusion is required to establish a sharp
interface in the mRNA
profile, the diffusion constant cannot  be too large.
As Fig.~\ref{thresholdFigure}G demonstrates,
increasing the diffusion constant may result in smearing the interface.
A corresponding increase in the interaction strength, $k$, can compensate for the increased diffusion and regain the interface sharpness (Fig.~\ref{thresholdFigure}H).
We will quantify these observations below.}

{Diffusing miRNAs can find themselves in one of two very different regions.
In the miRNA-rich region (including the region to the right of
the point where the transcription
profiles are equal)
miRNA decay mainly via processes independent of their interaction with the target.
In this region our model boils down to a simple diffusion process accompanied by
linear decay. Such processes are characterized by a length scale, denoted by $\lambda$, which
essentially measures how far a miRNA can travel (due to diffusion) before being consumed (by
autonomous degradation). It is thus an increasing function of the diffusion constant $D$, but a decreasing function of the independent decay rate $\beta_\mu$.
On the other hand, in the mRNA-rich region,
a miRNA decays mainly via co-degradation with its target.
In this region miRNAs decay faster, and one
expects them to be able to travel
over much shorter distances than in the miRNA-rich region. In fact, diffusion in this region is characterized by
another, smaller, length scale, denoted by $\ell$, which again increases with $D$ but is now a decreasing function of the interaction strength, $k$.
Explicit expressions for the two length scales are given in the Supporting Information (Eqs.~\ref{lambdaEq} and~\ref{ellEq}).}

To obtain a sharp interface in the mRNA profile, miRNAs should be
able to travel from the miRNA-rich zone into the mRNA-rich zone. This
means that the first length-scale, $\lambda$, should be of the same
order as the tissue length. This, for example, can be achieved if
the diffusion constant $D$ is {\em large enough}. On the other hand,
{the vicinity of the interface} is governed by the other length
scale $\ell$. This length scale is what determines the `width' of
the interface, namely the number of cells which exhibit intermediate
levels of mRNA expression {(see blue box in
Fig.~\ref{thresholdFigure}C)}. A sharp interface therefore means a
small value of $\ell$ and one way to achieve a small value of $\ell$
is to make the diffusion constant $D$ {\em small enough}. These two
contradicting requirements on $D$ suggest that there might be an
intermediate range of values for the diffusion constant that allows
for a sharp interface, but also raises the suspicion that this range
may be very small and require some {\em fine tuning}. 
This , however, is not the case: the fact that $\lambda$ is strongly dependent
on $\beta_\mu$ (while $\ell$ does not depend on $\beta_\mu$ at all),
and that $\ell$ strongly depends on $k$ (while $\lambda$ does not)
{means} that the range of allowed values of $D$ {can} be set as
large as needed.

{In the Supporting Information we develop an approximate analytical expression for the mRNA profile in terms of the various parameters and the `input' profiles $\alpha_m(x)$ and $\alpha_\mu(x)$.
There are two lessons to be learned from this {exercise}. {First,
the interface established by the mRNA-miRNA interaction is effectively impermeable to miRNA diffusion in the strong-interaction limit.}
The system thus separates into two parts which---in steady-state---do not exchange molecules between them. This property allows one to calculate the position of the sharp interface in the mRNA profile. }

{The second lesson  comes from the resulting equation for the interface position. This equation takes the form of a weighted spatial average of the difference between the two transcription profiles (Eq.~\ref{x0Eq}). In other words, in order to determine the position of the interface, one needs to sum over many nuclei. {In the sum}, each nucleus contributes the difference between the local transcription rates of the two RNA species. Clearly, these rates may be influenced by many factors, and in a description which is somewhat closer to reality
one would expect this difference to be  fluctuating around $\alpha_\mu(x)-\alpha_m(x)$.
However, the interface position is a sum of these fluctuating objects, and one might hope that the sum of these
fluctuations---which are uncorrelated---would be close to zero. In this case, the interface position would be robust to fluctuations {of this type}. Indeed, a stochastic simulation of the model shows no change in the interface position (or structure), as compared with the deterministic model discussed so far (see Supporting Information for details of our simulations).}

{In passing let us note that the conditions required so far --- 
namely, strong interaction between the miRNA and its target and small $\ell$ --- may be reached by making the mRNA
completely stable ($\beta_m \to 0$). However, our analysis shows that in this case the system would
never relax to a steady state, since target mRNAs would accumulate at the left end of the tissue without limit.} Our analysis here is therefore only applicable if the mRNA molecules undergo
autonomous degradation, in addition to the miRNA-dependent degradation.

\subsection{microRNAs can define sharp stripes of gene expression.}

\begin{figure*}
\includegraphics{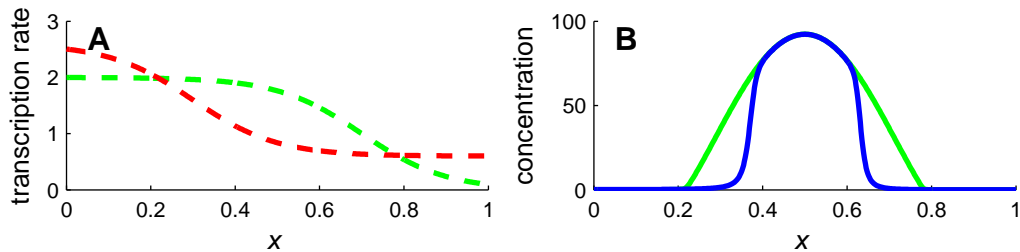}
\caption{Stripe of gene expression defined by microRNA interaction. {\bf A.} Transcription profiles of a miRNA (red) and its target (green), favorable for stripe generation. 
See Supporting Information for parameter values. {\bf B.} Expression of the target is restricted to the middle of the tissue even when the miRNA cannot diffuse ($D=0$, green). Mobility of the miRNA makes target expression more focused ($D=0.001$, blue): Target expression in the stripe is enhanced, and the stripe boundaries become sharp. 
\label{f.stripe}}
\end{figure*}

Using the insight gained in the previous section we briefly show how
a stripe is formed when the miRNA transcription profile
$\alpha_\mu(x)$ is similar to $\alpha_m(x)$ but displaced from it
(Fig.~\ref{f.stripe}A). Suppose, for example, that the synthesis of
an miRNA and its target are activated by the same transcription
factor. In any given nucleus, the two promoters experience the same
concentration of this transcription factor. However, they need not
react in the same way: if the binding affinity of one promoter is
stronger than that of the other, there will be intermediate
concentrations of the transcription factor where the first promoter
will be activated while the other will not. Such a scenario is
depicted in Fig.~2A, where a common transcription factor, which
exhibits a spatial gradient, activates the target gene as well as
the miRNA gene. In this case, the target promoter has higher
affinity to the  transcription factor than the miRNA promoter. Thus,
some cells in the middle of the developing tissue express {the target
mRNA} but not the {miRNA}.

{Unlike the case studied in the previous section, where the transcription profiles crossed at one point, here} the transcription profiles cross at two points.
%
{Let us retrace our steps in the previous section, by first considering the case of no diffusion.} For low values of the {interaction} rate $k$, the miRNA and mRNA profiles
are qualitatively similar to their transcription profiles.
As $k$ is increased however,
miRNA deplete mRNA levels
{at any position}
where $\alpha_\mu>\alpha_m$ and thus confine mRNA
expression to a stripe between the two crossing-points of the transcription profiles (green curve in Fig.~\ref{f.stripe}B).  
{Can diffusion make this profile sharper, as in the previous case?}
Indeed, {diffusing} miRNAs which survive annihilation
on the left and right diffuse into the interval between the two crossing points, and
establish sharp interfaces in the mRNA concentration profile. The resulting stripe resides
within this interval, but is narrower  (blue curve in Fig.~\ref{f.stripe}B).  
It is therefore important that parameters would allow for sharp
interfaces, without making the stripe too narrow (or even disappear). Therefore,
to sustain a well-defined stripe of gene expression, $\ell$,
which determines the interface width, must be much smaller than the distance between the two
crossing points of the transcription profiles.

{One can use the same analytic method mentioned
earlier to calculate the new positions of the stripe boundaries (see Supporting Information).
This exemplifies how the method can be used to analyze geometries of increasing complexity. }

\subsection{Standard experimental methods may be used to probe the sharpening mechanism.}

{The sharp interface that we predict can be detected directly in an imaging experiment,
provided the light intensity varies linearly with mRNA concentration, and that the spatial resolution is high enough.  However an experimental readout is often not so 
faithful to the underlying concentration profile. In the worst case, 
the apparatus' readout is binary:
concentrations below 
an apparatus-dependent threshold are not detected,
while concentrations larger than this
illicit a concentration-independent fluorescence intensity.
In such scenarios, it is impossible to tell a smooth and sharp 
concentration profile 
apart as both yield a sharp interface in the binary readout (Figs.~
\ref{cloneFigure}A-C).} In contrast, low spatial resolution may make a sharp 
interface appear smoother then it really is.   

{Fortunately, our model of miRNA-mediated morphogenic regulation possesses 
another signature that is visible at 
such coarse experimental resolution. To detect this signature, one needs to 
overexpress the miRNA in a small patch of cells (hereafter 
denoted the `patch'). 
Our model then predicts that this patch has a qualitatively different 
effect depending on which side of the interface it occurs. We confine ourselves in this section to the case of opposing transcription profiles of the miRNA and its target..

The technique one uses to {generate the patch}
 may differ according to the 
stage of development under consideration. 
In the early blastoderm stages of {\sl Drosophila} development,
for example, 
a Gal4 driver may be used to drive expression of the miRNA 
in those cells where an endogenous gene is expressed \cite{Blair2003}. 
Many endogenous genes are 
expressed in stripes along the anterior-posterior axis during these 
stages and some have dedicated enhancers for single stripes 
\cite{Small1992,Janssens2006}. {As an example, the yeast 
FLP-FRT recombination system has been used to misexpress 
the gap gene {\it knirps} in a stripe 
by placing it under the control of the {\it eve}
stripe 2 enhancer \cite{kosmanSmall1997}.}

In later stages of {\sl Drosophila} development, e.g.
imaginal discs, one technique is 
the random generation of {a mosaic of} mutant clones 
{(patches)}
by mitotic recombination \cite{Blair2003,Eldar2005}. 
The {patches} are generated at a low rate 
and one then screens for those embryos containing 
a single {patch}.

{We model the {localized overexpression of miRNA}
 by an effective {\em local} increase of the transcription rate
by an amount $\alpha_c=5$.  This increase in transcription rate occurs {in a small number of cells, which in our model is about $5\%$ }of the {tissue length}. More specifically, we choose a position $x_c$ for the center of the patch and, for every point $x$ that resides within a distance $w/2$ of $x_c$, we change the transcription rate from $\alpha_\mu(x)$ to $\alpha_\mu(x)+\alpha_c$. Here $w$ is the `width' of the patch, which takes a value {$w \simeq 5\%$ of the tissue length.}}

Consider positioning the patch first 
in the miRNA-rich region of the developing tissue, 
{ Fig.~\ref{cloneFigure}D.}
One sees that, even if positioned at a distance from the expression domain of
the target, the effect of the additional miRNA is to 
push the interface {toward the left.}
{The localized patch of cells}
therefore has a nonlocal effect.

{
Intuitively, one identifies three contributors to this effect.
First, of course, is the mobility of the miRNA. Diffusing miRNAs are the carriers of
information {from the patch to the other}
cells. However, these miRNAs need to be able to travel appreciable distance before they
are consumed, and thus the second factor is the fact that the {patch} was introduced
in a region where miRNAs only experience {autonomous} degradation, and can thus travel a distance
$\lambda$ (which {we have assumed to be} of the same order as the tissue length). 
Finally, the spreading {miRNAs} become localized again near the sharp interface {with}
the mRNA-rich region. This is simply due to the fact that miRNA cannot penetrate this region
over a distance that is larger than $\ell$ {(which we have assumed to be 
smaller than the width of a single cell).}
This intuitive picture is quantified in the Supporting Information. 
}

This experiment should be contrasted with one in which the
overexpressing patch is positioned in the mRNA-rich region, as shown in 
{Fig.~\ref{cloneFigure}E}. {Such ectopic miRNA expression
has a local effect}, with excess miRNA creating a {\em hole} 
in the mRNA expression domain. The hole edges constitute two 
additional interfaces in the system, the sharpness of each again  
determined by $\ell$. 




\begin{figure*}
\includegraphics{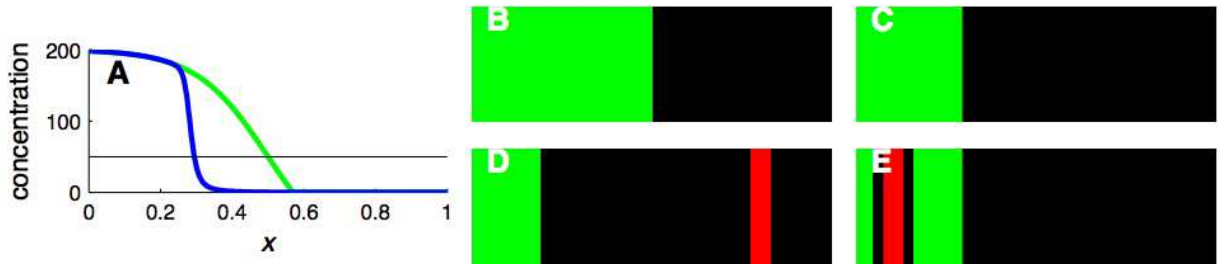} 
\caption{
Indirect test of the sharpening mechanism.
{\bf A.} Concentration profiles without diffusion of miRNA (green)
and with diffusion (blue). The black line denotes 
an apparatus-dependent threshold concentration. 
{\bf B, C.}  The readout of from an apparatus which amplifies any signal 
above its detection threshold  does not
distinguish between a smooth drop in mRNA levels 
(as occurs in B) from 
a sharp drop (the underlying profile in C).  
{\bf D.} A patch (red)
in the miRNA-rich region (posterior to the interface) 
pushes the interface to the left in a threshold assay.  
{\bf E.} A patch
in the mRNA-rich region (anterior to the interface) 
leaves the interface unaffected. 
See Supporting Information for parameter values.
\label{cloneFigure}}
\end{figure*}

\begin{figure}
\includegraphics{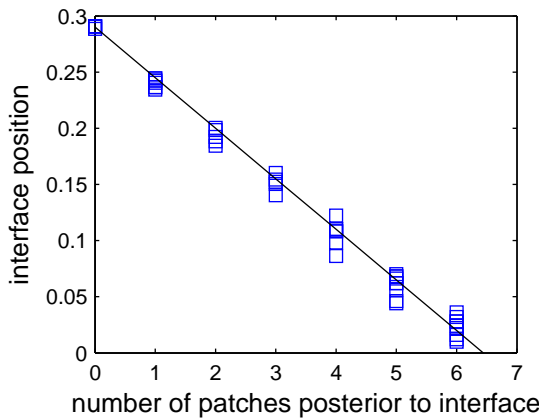}
\caption{
Quantitative test of the sharpening mechanism.
Simulated experiment in which the interface position $x_t$ is
measured in an ensemble of embryos having various numbers of patches. 
The theory predicts that  the interface displacement should 
be linearly proportional to the number of patches. 
\label{patchesFigure}}
\end{figure}

{
One can {go} further {and} make a quantitative prediction, relating the number of patches in a mosaic {of patches} with the lateral shift in the interface position. To first approximation, one needs to count the number of patches in the miRNA-rich region, and disregard completely the patches in the mRNA-rich region. The displacement of the interface position is then {linearly} proportional to this number; see Supporting Information for details. Simulated experimental results that would verify this prediction are shown in {Fig.~\ref{patchesFigure}.}
}

The distinct
nonlocal effect described above 
does not occur when the miRNA are unable to move between
cells.
Also, we have checked (for the parameters used 
in this study) that it does not occur when the miRNA acts
purely catalytically. Rather, {\em both} miRNA mobility {\em and} 
a strong interaction between miRNA and target are
required. The presence or absence of the nonlocal effect would therefore
confirm or falsify the hypotheses that miRNA are mobile and that 
they interact stoichiometrically 
with mRNA while in this mobile state.

\subsection{Irreversible localization of miRNA}

{Finally, we consider the case where the miRNA transcription profile overlaps that of its target.
This case is somewhat puzzling, as one would hardly see the point of going through this futile cycle.
In particular,} in the framework we have presented so far,
such a coincidence of transcription profiles would reduce
steady-state mRNA (and miRNA) levels
to negligible amounts in all cells.
{It is therefore not surprising, that this scenario is very atypical \cite{Farh12162005}. Where it does occur, e,g, possibly in the Hox gene network \cite{Mansfield2004}, one explanation may be that this is only a transient state.
In such a scenario, for example, the miRNAs keep the target mRNA from being expressed, until such time when their translation is required. At that time, transcription of miRNA may be stopped, or alternatively their interaction with mRNA may be inhibited. Induction of the target
{would then}  proceed rapidly. An example
{where}
miRNA expression {is regulated during the course of
 cellular development is} found in T-lymphocytes \cite{Neilson2007};
and regulation of miRNA-target interaction has been demonstrated in
neurons of {\it Drosophila} \cite{Ashraf2006}. Nevertheless, we
consider here the case where the overlapping transcription profiles
do represent---at least for some period of time---a steady-state,
and {show} that with a slight change in the dynamics, this layout
can still spatially separate the tissue into a region of high mRNA
concentration and a region where target mRNAs do not accumulate.}

 {The key ingredient required to rescue}
mRNA levels
 from global obliteration %
 is
the possibility that
only those miRNA localized to a cell interact with their target.
{Moreover,
miRNA} localization needs to be
irreversible; the case of rapid reversible localization
reduces to the model studied in previous sections.

The separation of transport and interaction may occur in a number
of ways. miRNA may be
transported from cell to cell in a vessel that needs to be
dismantled before miRNA can interact with mRNA.
Alternatively, the interacting form of the RISC complex may be
immobilized. Another possibility is that interaction between miRNA
and mRNA is limited to specific loci in the cell, from which miRNA
cannot escape.
These scenarios---and others---are summarized by a
model where miRNA acquire a mobile state immediately after synthesis,
and can switch into an interacting localized state at later
times
\beq
\mu \stackrel{q}{\longrightarrow} \mu_\ell .
\eeq
{This model, a straightforward generalization of \eqref{basicModel}, is given in
Eq.~\ref{herbie} of the Supporting Information, where we also solve it exactly. }

{Diffusion of the mobile miRNAs now occurs independently of the target,
since mobile miRNAs do not interact with the mRNA species. Accordingly, a steady-state
mobile miRNA profile is formed. This profile is smoother than the miRNA transcription profile
(since this profile is formed by diffusion with no interaction). In particular, it is lower in
those regions where the transcription profile $\alpha$
was {\em higher than average} and {higher} in
those where $\alpha$ was {\em lower than average}; See Fig.\ref{localisationFigure}A. }

{The steady-state mobile mRNA profile is the source for localized miRNA,
{which are thus} generated in a spatially dependent manner. Their generation rate
follows this profile, up to a factor $q$, the localization rate. This {\em generation rate profile} replaces the miRNA transcription profile of the previous sections. Thus, the role of the mobile miRNA is simply to set this profile. Left only with {the} localized miRNA and their mRNA targets,
diffusion should not be considered any further. This problem has already been discussed in the context of opposing transcription profile, and from that discussion we realize that mRNA concentration
still vanishes at any point where the miRNA generation rate profile exceeds that of the mRNA. In contrast, mRNA level is proportional to the difference between the two profiles wherever
{the mRNA transcription rate is larger}.
Fig.~5B shows how the tissue is separated into two regions.
{We have shown}
that this scenario boils down to a diffusion-free problem,
{so} the mechanism which could sharpen the
interface between the two regions is absent. As expected, the interface in Fig.~5B is smooth. }

 {The {value acquired by the} localization rate $q$ is important for the success of the scheme described above. If this rate
 is too high, the resulting mobile-miRNA profile -- and therefore, the localized miRNA generation profile -- would coincide with the transcription profile,
ultimately resulting in mRNA depletion in all cells. On the other hand, if the localization rate $q$ is too low, the generation rate of mobile miRNA may be lower than the mRNA synthesis rate in all cells, and have very little effect on their accumulation. Taken together, we find that for `optimal' performance, $q$ should be set around the miRNA degradation rate, $\beta_\mu$. However, our analysis shows, that no {\em fine-tuning} is required -- in fact, setting $q$ within an order of magnitude from $\beta_\mu$ would still yield two distinguishable regions along the developing tissue (Fig.~\ref{qFigure} and text in the Supporting Information).}

\begin{figure}
\includegraphics{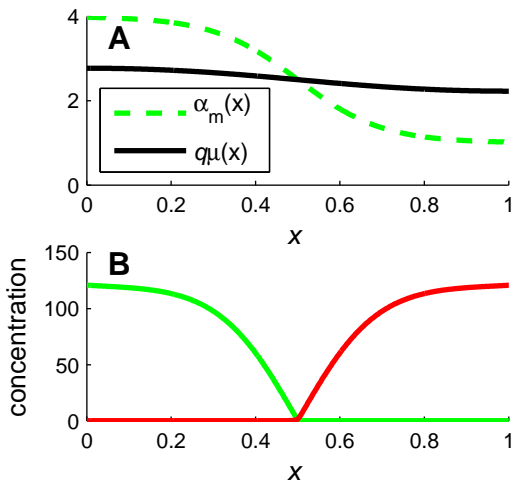}
\caption{Irreversible localization rescues the target from annihilation
throughout the tissue and sharpens its expression pattern.
{\bf A.} Steady state concentration profile of mobile miRNA (black) is proportional to the steady-state profile of the mobile miRNA,
is a broadened form of the transcription profile $\alpha(x)$ (green). In this figure $q=\beta_s$. {\bf B.} As a result, mRNA (green) accumulates only at the left side of the tissue.
The mRNA profile decays smoothly, and does not exhibit a sharp interface. 
\label{localisationFigure}}
\end{figure}

\section{Discussion}


In this study we have {analyzed} a model where miRNAs sharpen
target gene expression patterns by generating an
interface between high and low target expression.
Such an interface can be directly detected provided the signal is not
amplified to saturation and provided
the threshold for detection is sufficiently
low.

{A necessary condition for the interface to occur is that the
miRNA and target are co-degraded; a mode of miRNA-target
interaction where miRNAs only promotes mRNA degradation without being affected is insufficient.}
One can in principle test the {existence of
coupled miRNA-mRNA degradation by}
inhibiting the transcription of mRNA and monitoring its decay rate, which in this case would be time-dependent,  $\beta_{\rm eff}(t) =  \beta_m + ks(t)$. 
We note also that a
stoichiometric interaction may complicate the interpretation of sensor
transgene experiments \cite{Mansfield2004}, as the transgene
would then sequester miRNA and thereby alter the  original
expression patterns.



The interface between low and high mRNA levels
is characterized by low copy numbers of both RNA species. In such cases fluctuations in the copy number of either species may have macroscopic effects. For example, a small RNA-target pair in bacteria show enhanced fluctuations when their transcription rates become comparable \cite{Elf2003,Levine2007_noise}.  These fluctuations can in turn give rise to noise-induced bistability, which manifests itself experimentally as diversity in a population of cells \cite{Levine2007_noise}.
We performed Monte-Carlo simulations of our model, but found
that fluctuations have no macroscopic effect, even near the transition point
where copy numbers of both species are low.
This {in-built robustness to fluctuations}
arises because the interface position
is determined by an integrated transcriptional flux
which averages out individual cellular fluxes. 
Thus spatial averaging results in high spatial precision without smearing the interface.

The model studied here is defined in one spatial dimension, which we identified as the axis $x$
perpendicular to the interface between domains of gene expression. A two-dimensional extension of the model where the transcription profiles are independent of the $y$ axis and diffusion occurs
equally in both directions leaves the solution independent of the $y$ axis. 
However, we have further generalized the model to allow the mRNA transcription profile to have a slight dependence on $y$  (see Supporting Information).  We find that the results described above hold also in this two-dimensional case. Furthermore, we find that fluctuations in the interface position along the $y$-axis are considerably reduced by the miRNA regulation. A detailed analysis is deferred to a future study.

{Strong cooperative activation, as
often occurs in morphogenetic regulation at the transcriptional
level (e.g. Bicoid has about five binding sites
in target promoters of {\sl Drosophila}),
would seem to make pattern formation
by morphogens inherently susceptible
to temperature variations \cite{Houchmandzadeh2002,Segel1975}.
Nevertheless, embryonic patterning appears to be quite
robust to temperature variations, as has been documented for
Hunchback \cite{Houchmandzadeh2002} and for Eve \cite{Lucchetta2005}
in {\sl Drosophila}.
The only cooperative reaction required in the model
presented in this work is coupled degradation of
miRNA and mRNA, suggesting the possibility that miRNAs
filter fluctuations arising from temperature variations.}

Candidate systems in which to test the ideas put forth in
this study include the establishment of dorsoventral (adaxial/abaxial) leaf polarity in plants, 
as well as the segmentation of the early {\sl Drosophila} embryo. We now discuss these two
systems in some detail. 

\begin{figure}
\includegraphics{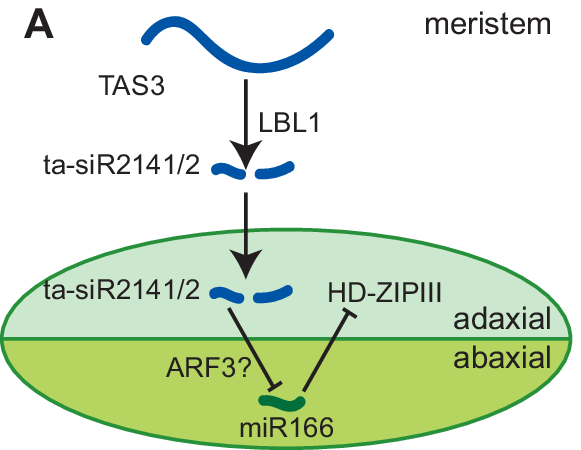}
\includegraphics{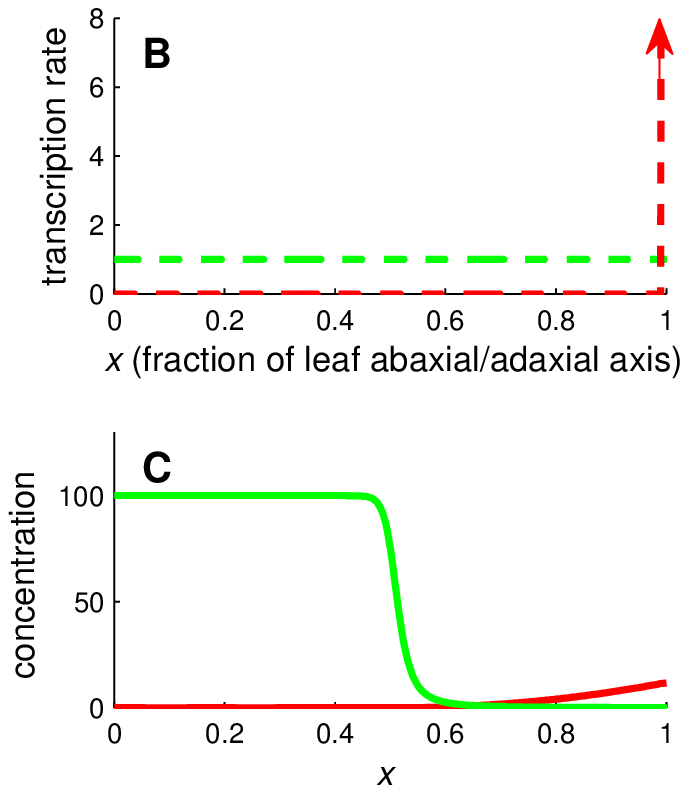}
\caption{Model for leaf polarity in maize. {\bf A.} In this model, TAS3 is processed in the meristem into several ta-siRNAs. Two of those, ta-siR2141/2, migrate into the adaxial side of the primordial leaf, where it inhibits miR166 (either directly or via ARF3). miR166 goes on to repress expression of HD-ZIPIII genes, which define adaxial fate.  {\bf B.} Transcription profiles. Transcription of ta-siR2141/2 (red) occurs in the shoot meristem, next to the adaxial side of the developing leaf. The target, either the miR166 or {\sl arf3}, is transcribed uniformly throughout the leaf (green). {\bf C.} Steady-state concentrations of ta-siR2141/2 (red) and its target (green). Diffusion of ta-siR2141/2 into the leaf restricts the expression of the target to the abaxial side, and -- assuming a non-catalytic interaction --  creates a sharp interface between the abaxial and adaxial domains. 
\label{leafFig}}
\end{figure}

Leaf polarity in plants is established shortly after the emergence of the leaf primordium from the meristem. Specification of leaf polarity depends on the Sussex signal \cite{Sussex51}, a meristem-borne signal that specifies adaxial cell fates. 
Members of class III of the homeodomain-leucine zipper (HD-ZIPIII) proteins specify adaxial fate \cite{Emery03,Juarez04}. In {\sl Arabidopsis} and in maize, the polar expression pattern of these genes results from their inhibition by two miRNAs, miR165/166, which exhibit a complementary expression pattern \cite{Juarez04,Kidner04}. Recently it has been shown that in maize, restriction of miR165/166 to the abaxial side of the developing leaf depends on the polarized expression of LBL-1, a protein involved in the biosynthesis of trans-acting RNAs, ta-siR2141/2 \cite{Nogueira07}. Possible targets of ta-siR2141/2 include members of the {\sl arf3} gene family (a transcription factor that is expressed abaxially) as well as members of the miR166 family \cite{Nogueira07}. 
While miRNAs in plants are thought to act mainly cell-autonomously \cite{Alvarez06}, DCL4-dependent siRNAs, such as ta-siR2141/2, may exhibit cell-to-cell movement \cite{Voinnet05}. 

The following model is consistent with these data (Fig.~\ref{leafFig}A). The RNA transcript {\sl TAS3} is cleaved to produce ta-siR2141/2 in the meristem. These small RNAs  then propagate (diffuse) into the adaxial side of the leaf, inhibiting the expression of miR166 either directly or through the ARF3 transcription factor. The target (either miR166 or ARF3) is transcribed uniformly throughout the leaf, and is localized to the cell where it is synthesized.  If one further assumes that the interaction between ta-siR2141/2 and its target is non-catalytic, then this model belongs to the class of models 
studied in this work, and can therefore exhibit a sharp interface between the abaxial domain of high target expression  and the adaxial domain of no expression; see Fig.~\ref{leafFig}. 
In agreement with this model is the low abundance of ta-siR2141/2 in {\sl Arabidopsis} \cite{Allen05,Williams05},  despite their distinct phenotypic role. 

Early embryonic development in {\sl Drosophila} proceeds via a
cascade of gene activities that progressively refine expression
patterns along the anterior-posterior axis of the embryo. A recent
study of the expression patterns of nascent miRNA transcripts
suggests that a number of miRNAs may play a role in this process.
The miRNAs miR-309clus, miR-10 and iab-4
(which all
reside between annotated mRNA genes on the genome), and
miR-11, miR-274 and miR-281clus (which all reside  within introns
of annotated genes) are {all} expressed in
a graded fashion along
the anterior-posterior axis of the blastoderm embryo
\cite{Aboobaker2005,Kosman2004}.

The complementary transcription profiles of iab-4 and its target Ubx
at stage 5 {of development}
make this miRNA-target system
a candidate for the sharpening mechanism proposed in this study.
{The early Ubx transcript pattern is broadly distributed
over the posterior half of the embryo, becoming localized to a stripe
at the centre of the embryo by the completion of cell formation
\cite{AkamMartinezArias1985,Ronshaugen2005},
probably as a result of transcriptional repression by
Hunchback in the anterior and posterior regions of the embryo
\cite{ZhangBienz1992}. The nascent transcript profile
of its regulator, iab-4, is broadly distributed posterior to this
stripe \cite{Ronshaugen2005}.}

{It may be the case that iab-4 is also expressed before cell
formation and that the absence of cell walls makes iab-4 mobile.
The much larger Ubx mRNA, on the other hand, may be effectively
stationary on the timescales of interest \cite{Gregor2007b}.
Furthermore, the transcription profiles of iab-4 and Ubx at stage 5 do not
seem to overlap \cite{Ronshaugen2005}, suggesting that iab-4 intercellular
mobility may be crucial to allow it to interact with Ubx at this
stage of development. Assuming then that only the miRNA iab-4 is mobile,
the complementary expression patterns of iab-4 and its
target Ubx measured in Ref.~\cite{Ronshaugen2005}
is consistent with our model of miRNA-induced sharpening. Sample
profiles predicted by the model are shown in Fig.~\ref{ubxDiscussion}.}

\begin{figure}
\includegraphics{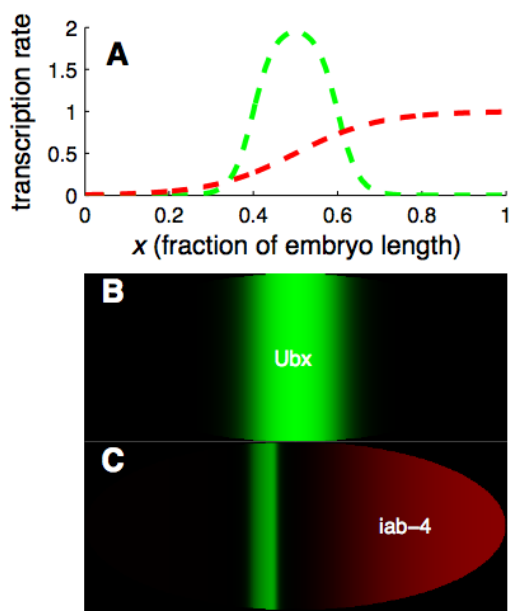}
\caption{Model for sharpening of Ubx expression boundaries by iab-4. {\bf A.} The assumed transcription profiles of Ubx (green)
and iab-4 (red).{\bf  B.} In the absence of iab-4 (or
when the interaction between iab-4 and Ubx is inhibited), the
expression pattern of Ubx is broad and smeared. {\bf C.} When iab-4
and Ubx interact, the expression pattern of Ubx becomes narrow and sharp.
The diffusion constant is $D=0.1$ and the co-degradation rate is $k=1$. 
The embryo-like shape is for illustrative purpose only.
\label{ubxDiscussion}}
\end{figure}

{A possible difficulty with regard to applying our
model to Ubx/iab-4 is that
the system may not have reached steady state before stage 6 when
cells begin to migrate.} In particular,
no Ubx protein was detected at stage 5, possibly because
of the time needed to transcribe the large Ubx locus \cite{Ronshaugen2005}.

Like iab-4, the miRNA miR-10 is also expressed at stage 5
in a broad
posterior region along the anterior-posterior axis \cite{Aboobaker2005}.
The homeotic gene Scr is a predicted target of miR-10 \cite{Enright2003}
and is also expressed in the blastoderm at stage 5 \cite{Gorman1995}.
The miR-10 site in the Scr 3'UTR is likely to be functional because
the pairing is well conserved
in all drosophilid genomes and because the miRNA site is conserved in the Scr
genes in mosquito, the flour beetle and the silk moth \cite{Brennecke2005}.
Unlike Ubx, the protein of Scr is detected at this stage of development in
a stripe of ectodermal cells about 4 cells wide in the parasegment-2
region, though it may not be functional at this time as
the protein (a transcription factor)
was not localized to the nucleus \cite{Gorman1995}.
This spatial expression pattern
is proximal to the anterior limit of miR-10 expression
\cite{Gorman1995,Aboobaker2005}. Hence
the interaction of miR-10 with Scr at stage
5 of {\sl Drosophila} development is also a
candidate for the sharpening mechanism.


The sharpening mechanism is most effective when the spatial transcription
profiles of miRNA and target are regulated in such a
way as to be mutually exclusive.
The genomic locations of the miRNAs iab-4 and miR-10 are proximal
to their targets, {which is certainly}
consistent with the possibility of
coordinated regulation \cite{Pearson2005}.



\section{acknowledgments}
It is a pleasure to thank William McGinnis, Martin F. Yanofsky
and Jeffrey A. Long for useful discussions.
This work has been supported in part by the NSF-sponsored Center
for Theoretical Biological Physics (grant numbers PHY-0216576
and PHY-0225630).

\cleardoublepage
\onecolumngrid
\section{Supporting Information}
\hspace{2.0cm}
\twocolumngrid
\setcounter{equation}{0}
\setcounter{page}{1}
\setcounter{figure}{0}

 \renewcommand{\theequation}{S\arabic{equation}}
 \renewcommand{\thefigure}{S\arabic{figure}}
\subsection{Parameter Values}
Throughout, $\beta_m=\beta_\mu=0.01$, $k=1$. The transcription profiles of Fig.~\ref{thresholdFigure} are
\beqa
\label{tsxprofiles}
\alpha_m(x) & = & 0.5 A_m [ \tanh((x_{tsx}-x)/\lambda_{tsx}) + 1 ]\\
\alpha_\mu(x) & = & 0.5A_\mu [ \tanh((x-x_{tsx})/\lambda_{tsx}) + 1 ],\nonumber
\eeqa
where $A_m = 2$, $A_\mu = 1$, $x_{tsx} = 0.5$ and $\lambda_{tsx}=0.2$. In the stripe geometry (Fig.~\ref{f.stripe}), 
the transcription profile for $m$ is as above, with $x_{tsx}=0.3$. The transcription profile of the miRNA,  is given by 
\beq
\alpha_\mu(x)  = A_{\mu0}+0.5A_\mu [ \tanh((x_{tsx}-x)/\lambda_{tsx}) + 1 ],
\eeq
with $A_\mu=2, A_{\mu0}=0.6$ and $x_{tsx}=0.7$. To test our model for irreversible localization 
we use the same transcription profiles as in \eqref{tsxprofiles}, with $A_m=A_\mu=1$. 

In the discussion we outline possible applications in two systems: leaf polarity in maize and segmentation in the early {\sl Drosophila} embryo. Here we did not aim to estimate parameters from
experimental data (which, in most cases, is not quantitative enough to allow for parameter inference).
Instead, parameters were chosen arbitrarily to allow clear demonstration of possible results. 
In the case of leaf polarity (Fig.~\ref{leafFig}) we choose $\alpha_m(x)=A_m, \alpha_s(x)=A_\mu\theta(x-x_{tsx})$ with $A_m=1, A_\mu=50$ and $x_{tsx}=0.99$. Here $\theta(x)$ is the unit step function.  In the {\sl Drosophila} embryo (Fig.~\ref{ubxDiscussion}), the transcription profile of the iab-4 miRNA is the same as in 
\eqref{tsxprofiles}, while the Ubx mRNA transcription profile is given by 
\beq
\alpha_m(x) = 0.5 A_m [ \tanh(|x_{tsx}-x|/\lambda_{tsx}) + 1 ]\;.
\eeq

To study a two-dimensional extension of this model, we let the mRNA transcription profile have a slight dependence on the $y$-direction, 
{\it e.g.} by setting $x_{tsx}$ in $\alpha_m(x)$ (Eq.~\ref{tsxprofiles}), to $x_{tsx}=\sin(4\pi y)/10$.

\subsection{{Strong-interaction limit}}
{Here we define, for the case of no diffusion, what we mean by the
strong-interaction limit, but the limit is equally valid when
diffusion is present.} We make the critical assumption that at any
cellular position the decay of the minority species is dominated by
{coupled degradation}; by rescaling Eq.~\ref{annModelEqs} of the
main text, this condition amounts to \beq \mbox{max}\left\{
\frac{k\alpha_m(x)/\beta_m}{\beta_\mu},
\frac{k\alpha_\mu(x)/\beta_\mu}{\beta_m}\right\} \gg 1, \mbox{ for
all $x$} . \label{criticalAssumption} \eeq In the absence of
diffusion,  Eq.~\ref{annModelEqs} boils down to a quadratic equation
for the steady-state mRNA level $m$, and one can immediately write
the solution \beq m = \beta_m^{-1}\left[\alpha_m-\alpha_\mu(1+
\epsilon)+\sqrt{[\alpha_m+\alpha_\mu(1+
\epsilon)]^2-4\alpha_m\alpha_\mu}\right]\;, \eeq where
$\epsilon\equiv\beta_m\beta_\mu/(k\alpha_\mu)$, which is assumed in
\eqref{criticalAssumption} to be small. To zeroth order in
$\epsilon$, this expression simplifies to
 \beq m \approx
\frac{[\alpha_m-\alpha_\mu]_+}{\beta_m} , \label{thresholdResponse}
\eeq where $[x]_+=\mbox{max}\{0,x\}$,
as depicted in Fig.~\ref{thresholdFigure}C.

\subsection{Analytical Approximation}

To understand the origin of the length scales {$\lambda$
and $\ell$}, and their
relation to the tissue length $L$ when the interface is sharp,
consider first
the region of space where miRNA are in the
majority. In this region, where $k\mu \gg \beta_m$,
we neglect the autonomous-degradation term in \eqref{annModelEq1},
yielding $\alpha_m  = km\mu$ and thus
\beq
0  =  \alpha_\mu -\alpha_m - \beta_\mu\mu + D\mu''\;. \label{miRNAEq}
\eeq Hence miRNA
are produced at an effective rate $\alpha_\mu - \alpha_m$ and
diffuse over distances of order
\beq
\lambda =
\sqrt{\frac{D}{\beta_\mu}}\;,
\label{lambdaEq}
\eeq
which, {as we have argued in the main
text, should} be comparable to the
{tissue length, $\lambda\sim L$}.
On the other hand,  in the
mRNA-rich zone $km \gg \beta_\mu$ and so the only length scale
available to the miRNA is
\beq
\ell = \sqrt{\frac{D}{k\alpha_m^*/\beta_m}} \;,
\label{ellEq}
\eeq
where $\alpha_m^*$ is a typical value
of $\alpha_m(x)$ in the mRNA-rich zone.
For the mRNA this is the only length scale that competes
with the spatial layout provided by the transcription profile
{and so it must determine the interface
width up to a constant prefactor $p$. Hence the
second condition for a sharp interface is that $p\ell\ll L$. }
In agreement with the expression in Eq.~\ref{ellEq}, derived on
hueristic grounds, our numerical solutions show that
the interface {becomes broader
when the co-degradation rate $k$ is decreased (Fig.~\ref{thresholdFigure}G) or when
the diffusion constant $D$ is increased
(Fig.~\ref{thresholdFigure}H).}
{We note that the limit $p\ell\ll L$ is equivalent
to
\beq
k\alpha_m^*/\beta_m \gg D/L^2.
\label{lossDueToDiffusion}
\eeq In other words, a sharp interface
arises when the co-degradation rate of miRNA and target dominates
the rate of diffusion over macroscopic distances.}
Since we are neglecting diffusion of miRNA
in the mRNA-rich region,
the mRNA profile there is given again by \eqref{thresholdResponse}.

Microscopically, miRNA in the miRNA-rich zone diffuse in a landscape
dominated by {autonomous degradation},
leading to the decay length
$\lambda$ in \eqref{lambdaEq}. Upon entering the mRNA-rich
region, {co-degradation}
suddenly overwhelms {autonomous degradation of miRNA}
($km\gg\beta_\mu$), {and} the miRNA are
faced with an effective
absorbing boundary. We therefore expect the miRNA concentration to
vanish as one approaches the interface from
the right. In addition, our picture asserts that
the miRNA concentration is vanishingly small everywhere on the left of the
interface. Taken together, these two properties impose zero
miRNA concentration and zero miRNA {diffusive} flux at the interface.
These two boundary conditions on the miRNA dynamics at
the interface between mRNA-rich and miRNA-rich regions, together
with the zero-flux condition at $x=L$, allow us to determine the
position $x_t$ of the interface. Furthermore, the interface must lie
{in the region defined by $\alpha_m>\alpha_\mu$}
because {co-degradation} can dominate
{autonomous degradation of miRNA only if
there is a reservoir of mRNA to co-degrade} with.

Armed with this insight into the miRNA profile we may now solve
\eqref{miRNAEq},
subject to the boundary conditions $-D\mu'(x_t) =
-D\mu'(L) = 0$ and $\mu(x_t)=0$,
in terms of a Green's function. Making use of the zero-flux boundary
conditions, the Green's function of \eqref{miRNAEq} is \beq
g(x,s) = \left\{
\begin{array}{ll}
G(x,s) & \mbox{if $x<s$} \\
G(s,x) & \mbox{if $x>s$}
\end{array}
\right. , \label{GreensFunctionEq} \eeq where \beq \lambda \, G(x,s)
=
\frac{\cosh\left(\frac{x-x_t}{\lambda}\right)\cosh\left(\frac{L-s}{\lambda}\right)}{\sinh\left(\frac{L-x_t}{\lambda}\right)}
. \eeq
The miRNA profile is then a weighted spatial average of the net
transcriptional flux of miRNA to the right of the interface
\beq
\beta_\mu \mu(x) = \int_{x_t}^L [\alpha_\mu(s)-\alpha_m(s)] g(x,s) \, ds .
\label{anasol}
\eeq
Employing the zero-concentration boundary condition
$\mu(x_t)=0$, we arrive at the following implicit equation for $x_t$
\beq
\int_{x_t}^L [\alpha_\mu(s)-\alpha_m(s)] \, g(x_t,s) \, ds = 0.
\label{x0Eq}
\eeq
Solving for $x_t$ requires
knowledge only of the
transcription profiles. Moreover,
one immediately sees that $x_t$ can tolerate
fluctuations in the transcription profiles which preserve the
integral. This should be contrasted with the non-diffusive
case {in which $x_t$ reduces to the crossing point
of the transcription profiles, $\alpha_m(x_t)=\alpha_\mu(x_t)$,
which is less robust to small-number fluctuations.}

\begin{figure}
\includegraphics{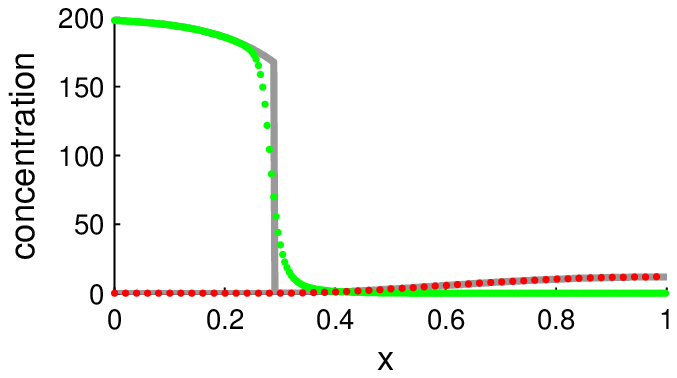}
\caption{A comparison of the exact numerical concentration profiles
with the analytical approximation (grey lines) described in the text.
Note that the interface is always sharp in the analytic approximation.
The diffusion constant here is $D=0.01$ and the 
miRNA-mRNA interaction parameter
is $k=1$. Autonomous degradation rates are 
$\beta_i = 0.01$. \label{compareAnalyticSupporting}}
\end{figure}

In Fig.~\ref{compareAnalyticSupporting} we compare the analytical
expressions in Eqs.~\ref{thresholdResponse}, \ref{anasol} and~\ref{x0Eq}
with the exact numerical
solution of \eqref{annModelEqs}.
{As expected the agreement is good because co-degradation
dominates both autonomous degradation
(Eq.~\ref{criticalAssumption})
and loss due to diffusion (Eq.~\ref{lossDueToDiffusion}).}

%

\subsection{Stochastic Simulations}

We used the Gillespie algorithm
to stochastically simulate the reaction and diffusion events
on a one-dimensional grid of cells \cite{Gillespie1977}.
In this algorithm the next event,
as well as the time to the next event, are chosen randomly.
A simulation using 100 cells 
(the approximate anterior-posterior length of the 
 {\sl Drosophila}
embryo during cycle 14) is compared with the 
corresponding deterministic solution in Fig.~\ref{stochSimFigure}.
Surprisingly, the deterministic 
solution is a good approximation 
for mRNA  anterior abundances  as low as 20 molecules per cell.
As expected, in cases where the predicted interface is of the order of a single cell
we find that the solution to our model underestimates the width of the interface. This, for 
example, is the case 
when the developing tissue becomes as small as
10 cells---the approximate size of the leaf-organ primordium 
during plant development.  

\begin{figure}
\begin{center}
\includegraphics{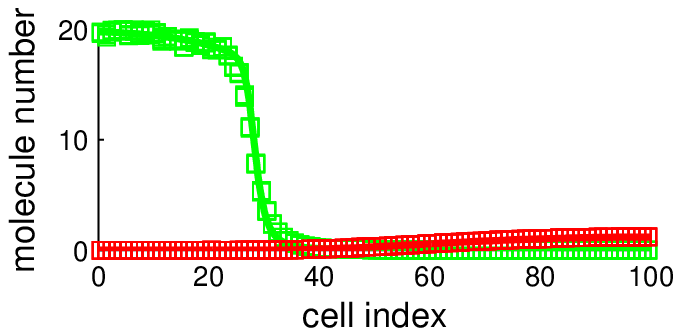}
\end{center}
\caption{Temporal averages (squares) of a stochastic simulation of the model,
compared with the corresponding mean-field solution (solid 
lines), in a developing field comprising 100 cells.
The diffusion constant is $D=1000$ and the 
miRNA-mRNA interaction parameter
is $k=100$. The autonmous degradation rates are 
$\beta_i = 0.1$. \label{stochSimFigure}}
\end{figure}

\subsection{Stripe Boundaries}

{We denote the left-most boundary of the stripe
by $x_{t1}$ and
the right-most by $x_{t2}$.} The
location of these interfaces is determined
analytically by solving \eqref{miRNAEq}
with zero-flux boundary conditions
in the interval $[0,x_{t1}]$ (and then enforcing $\mu(x_{t1})=0$) and
in the interval $[x_{t2},L]$ (and then enforcing  $\mu(x_{t2})=0$).
In the region between the interfaces the mRNA profile is approximately given
by $m = [\alpha_m-\alpha_\mu]/\beta_m$; in the portion of the
developing tissue complementary to
this the mRNA profile is negligible.
{The analytic profiles for $\mu$ and $m$ are compared with the
exact numerical solutions in Fig.~\ref{compareAnalyticStripeSupporting}.}

\begin{figure}
\includegraphics{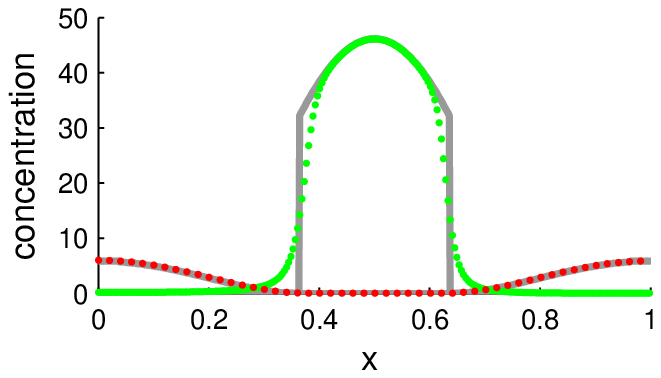}
\caption{A comparison of the exact numerical concentration profiles
with the analytical approximation (grey lines) described in the text
when the transcription profiles are aligned so as to produce a stripe.
See caption of Fig.~\ref{compareAnalyticSupporting} for 
parameter values. \label{compareAnalyticStripeSupporting}}
\end{figure}



\subsection{Experimental Prediction}

{The nonlocal effect predicted when the miRNA is
overexpressed in a small patch of cells can be understood
with the aid of \eqref{miRNAEq}. First, integrate this equation
from the interface to $L$, both with $(\hat{x}_t)$
and without $(x_t)$ the patch. Then,
neglecting the change in the miRNA concentration induced by the patch,}
one can show that
\beq
\alpha_c w
\approx
\int_{\hat{x}_t}^{x_t} \alpha_m .
\label{conservationLaw2}
\eeq
{In other words} miRNAs from the patch
diffuse across the tissue towards the region $[\hat{x}_t,x_t]$
and annihilate all mRNA produced there that would
otherwise have maintained the interface at $x_t$.

{As mentioned in the main text,}
we can make a quantitative testable prediction when there are a
number of independent patches. To see this recall
that Eq.~\ref{conservationLaw2} is an integral relationship.
Hence,
{in the corresponding equation for multiple patches,}
the left-hand side is proportional to the number of patches,
provided we make the reasonable assumption that the patches are
uniform in size and transcription rate.
Similarly, the right-hand side is proportional to the interface shift,
$x_t-\hat{x}_t$,
provided the mRNA transcription profile
is sufficiently flat in the interval $[\hat{x}_t,x_t]$.
Hence
the interface position $\hat{x}_t$ decreases
linearly with the number of {patches.}
This prediction
can readily be tested using an ensemble of embryos
with varying numbers of {patches}.
{Simulated experimental results that would verify this prediction are shown
in Fig.~\ref{patchesFigure} in the main text.}
\begin{figure}[t]
\includegraphics[width=8cm]{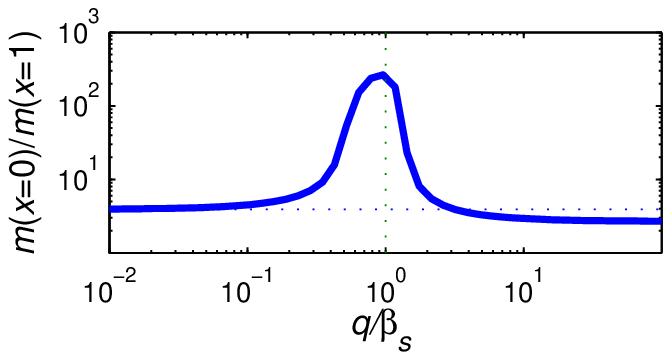}
\caption{The ratio between mRNA concentration at the left
($x=0$) and right ($x=1$) ends of the tissue, for the
transcription profiles and parameters of
Fig.~\ref{localisationFigure}. In the absence of miRNA regulation,
this ratio is $~\sim 3.9$ (horizontal dotted line). microRNA
regulation is most effective in clearing residual mRNA from `off'
cells when $q \simeq \beta_\mu$ (vertical dotted
line).\label{qFigure}}
\end{figure}

\subsection{Irreversible Localization}

We employ this flavor of the model for the special case that the
miRNA and mRNA transcription profiles coincide. Mobile miRNA can no
longer interact with their targets; only when they flip irreversibly
to the localized form \beq \mu \stackrel{q}{\longrightarrow}
\mu_\ell . \eeq can they do so. This adds a new equation to our
model which now takes the form
\begin{subequations}\label{herbie} \beqa \label{herbiea}
0 & = & \alpha - \beta_m m - k m \mu_\ell \\ \label{herbieb}
0 & = &  q \mu- \beta_\ell \mu_\ell - k m \mu_\ell \\\label{herbiec}
0 & = &  \alpha - \beta_\mu \mu  - q\mu + D
\mu''\;. \eeqa \end{subequations}
Here $\mu_\ell$ is the concentration of miRNA in the
localized state, and $q$ is an effective rate at which mobile miRNA
switch into the interacting localized mode. The common transcription
profiles of mobile miRNA and mRNA is denoted by $\alpha(x)$.
%
The
mobile-miRNA profile $\mu(x)$ is determined only by
the third equation and,
by analogy with Eq.~\ref{anasol}, is a weighted spatial
average of $\alpha$.
As such the effective transcription profile of immobile miRNA $q\mu(x)$
is reduced on the left and increased on the right relative
to that of the target $\alpha(x)$
%
The first two equations then reduce to the zero-dimensional case
considered earlier in which miRNA clear up mRNA to the right of the
point where $q\mu(x)$ and $\alpha(x)$ cross.
%

The value of $q$ which is most favorable for the failsafe function
is $q\approx\beta_\mu$. To understand why, it is helpful to consider
two limits. In the first, {localization} dominates
all other loss {processes} for mobile miRNA, i.e. $q\gg\beta_\mu$ and
$q\gg D/L^2$.
Hence all mobile miRNA immediately become immobile,
the transcription profiles of mRNA and immobile miRNA coincide
($\alpha \approx q\mu$,  Eq.~\ref{herbiec}), and all mRNA are immediately
degraded in all cells. In the second limit, {localization}
is negligible
compared with
{independent degradation},
$q\ll \beta_\mu$, and very few mobile miRNA become immobile.
Consequently $q\mu\ll \alpha $ and there are practically no immobile
miRNA available to affect the target. One measure of the efficacy of
the failsafe mechanism {(threshold response)} is the ratio between
mRNA concentration at the two ends of the tissue, $m(0)/m(L)$. With
the transcription profiles and parameters of
Fig.~\ref{localisationFigure}, this ratio would be $\sim 3.9$ in the
absence of miRNA regulation, and $\sim 257.4$ in its presence. In
Fig.~\ref{qFigure} we plot this ratio as a function of $q$, keeping
all other parameters as in Fig.~\ref{localisationFigure}. One sees
that the failsafe mechanism is most effective in the intermediate
regime where $q \approx \beta_\mu$.

\end{document}